\documentstyle[12pt]{article}
 \def\newpic#1{%
    \def\emline##1##2##3##4##5##6{%
       \put(##1,##2){\special{em:point #1##3}}%
       \put(##4,##5){\special{em:point #1##6}}%
       \special{em:line #1##3,#1##6}}}
 \newpic{}
\def\be{\begin{equation}}
\def\ee{\end{equation}}
\def\bea{\begin{eqnarray}}
\def\eea{\end{eqnarray}}
\def\text{\rm}
\def\be{\begin{equation}}
\def\ee{\end{equation}}
\def\bea{\begin{eqnarray}}
\def\eea{\end{eqnarray}}

\begin{document}
\begin{titlepage}

\begin{center}
{\LARGE $S$-matrix representation of the finite temperature\newline
propagator in $\lambda\varphi^4$-QFT}
\end{center}
\vspace{1cm}
\begin{center}
{\large A.I.~Bugrij, L.L.Jenkovszky, V.N.Shadura}
\footnote{E--mail: abugrij@gluk.apc.org}\\
\mbox{} \\
{\it Bogoliubov Institute for Theoretical Physics,} \\
{\it 14, Metrologichna str., Kiev 252143, Ukraine}
\end{center}

\vskip 1 cm
\begin{abstract}
The two-point Green function of the massive scalar
$(3+1)$-quantum field theory with $\lambda\varphi^4$ interaction at
finite temperature is evaluated up to the 2nd order of perturbation
theory. The everaging on the vacuum fluctuations is separated from the
everaging on the thermal fluctuations explicitly. As a result, the
temperature dependent part of the propagator is expressed through the
scattering amplitudes. The obtained expression is generalized for
higher orders of perturbation theory.
\end{abstract}

\end{titlepage}

\setcounter{page}{2}

\section{Introduction}

In 1969 R.~Dashen, S.~Ma and H.J.~Bernstein  have suggested [1] a
generalization  of the Beth--Uhlenbeck formula. They have
obtained the complete virial expansion for the grand potential
$\Omega$ where the $n$-th virial coefficient was expressed in terms of
special traces of the $n\to n$ $S$-matrix elements
\be
\beta(\Omega-\Omega_0)={1\over 4\pi
i}\sum_{n=2}^{\infty}z^n\int\limits_{n\cdot m}^{\infty}\, dE\,
e^{-\beta E} \biggl({\text Tr}_n\,
AS^{-1}{\stackrel{\leftrightarrow}{\partial}\over \partial
E}S\biggr)_c,
\ee
($T=1/\beta$ is the temperature, $z$ is the
activity, $S$ is the scattering matrix, $A$ is the exchange operator,
$m$ is the mass of a particle and $\Omega_0$ is the grand potential of
the free particle system). The compact and invariant form of the
eq.(1) was the reason  to  claim [1] the validity of this
representation also in the relativistic case. We share this point of
view in spite of absence of any direct quantum field theory
derivations.

In 70-th we exploited this formalism -- so called $S$-matrix
formulation of statistical mechanics -- for the phenomenological
investigation of the hot hadron matter thermodynamics [2--5]. The
equation of state derived on the basis of Regge phenomenology
for the scattering amplitudes appeared to be essentially nonideal at
GeVs temperatures. We have used it for  some
astrophysical and cosmological applications [6--8]. In particular due
to a Van-der-Waals-like nonideality of our equation of state we have
obtained in the framework of the Fridman model the exponential
expansion of the Universe in the vicinity of the phase transition,
providing for a solution the of horizon, flatness,
isotropy, primary fluctuations  and some others problems.
A few years later   Guth published [9]  his
outstanding Inflationary Universe scenario
based on a
quantum field theory approach.

At present  the general fashion for the theoretical treatment of
hadron matter (quark-gluon plasma) is ingenuous QCD: perturbation
theory, lattice simulations. Nevertheless we believe that the
$S$-matrix approach to the problem is fruitful at least in
a qualitative and heuristic sense (as for quantitative analysis see for
example ref.[10]). To be more acceptable $S$-matrix approach needs of
course in detailed QFT tests. The perturbation theory analysis is a
step in this direction.

In this work we carry out two-loop calculations for the temperature
dependent part of the 1PI two-point Green function of
$\lambda\varphi^4$ QFT -- the simplest but nontrivial quantum field
model.  We find that it can be represented through the
thermal everages of corresponding renormalized scattering amplitudes.

\section{The model}
\medskip

We considered a scalar field $\varphi(x)$ in a box with
the volume $V$ at the temperature $T=\beta^{-1}$. The points of the
Euclidian space-(imaginary)time are enumerated by coordinates
$x=(x_0,\vec x)$. The scalar product is
$$
(xy)=x_0y_0+(\vec x\vec y),\qquad x^2=x_0^2+\vec x\,^2.
$$
The field $\varphi(x)$ is periodical along the time direction
$\varphi(x_0,\vec x)=\varphi(x_0+\beta,\vec x)$ according to
Bose-Einstein statistics. Space boundary conditions are irrelevant in
thermodynamical limit $V\to \infty$.

The action is
$$
S[\varphi]=\int d^3x\int\limits_{0}^{\beta}dx_0\bigl({\cal
L}_0(\varphi)+{\cal L}_I(\varphi)\bigr)
$$
where the free-field Lagrangian is
\bea
{\cal
L}_0(\varphi)&=&{1\over2}\varphi(x)(m_0^2-\Box)\varphi(x),\nonumber\\
\Box&=&{\partial^2\over\partial x^2_0}+ {\partial^2\over\partial
x^2_1}+{\partial^2\over\partial x^2_2}+{\partial^2\over\partial
x^2_3}.\nonumber
\eea
The interaction due to
$$
{\cal L}_I(\varphi)={\lambda_0\over4!}\varphi^4(x),
$$
where $m_0$, $\lambda_0$ are the bare mass and the coupling constant.
The generating functional is $$
\exp\{Z[J]\}=\int D\varphi e^{-S[\varphi]-(J,\varphi)}, $$
where $J(x)$ is the external source and $$
(J,\varphi)=\int d^3x\int\limits_{0}^{\beta}dx_0J(x)\varphi(x).  $$
The Green functions are defined as usual [11,12] $$
G^{(n)}(x_1,\dots,x_n)={\delta^nZ[J]\over \delta J(x_1)\dots \delta
J(x_n)}.
$$
The one particle irreducible (1PI) Green functions are
$$
\Gamma^{(n)}(x_1,\dots,x_n)={\delta^nW[\psi]\over\delta\psi(x_1)\dots
\delta\psi(x_n)},
$$
where
$$
\psi(x)={\delta Z[J]\over\delta J(x)},\qquad {\delta
W[\psi]\over\delta \psi(x)}=-J(x).
$$
We use the Fourier transformation
(direct and inverse) in the appropriate normalization
\bea
f(x_0,\vec x)&=&{T\over(2\pi)^3}\sum_{p_0}\int
d^3pe^{-ip_0x_0-i\vec p\vec x}\widetilde{f}(p_0,\vec p),\nonumber\\
\widetilde{f}(p_0,\vec p)&=&\int
d^3x\int\limits_{0}^{\beta}dx_0e^{ip_0x_0+i\vec p\vec x}f(x_0,\vec
x),\nonumber\\
p_0&=&2\pi\ell T,\qquad \ell=0,\, \pm1,\,\pm2,\dots.\nonumber
\eea

\section{1PI $\bf{\Gamma^{(2)}}$-function}

At $T=0$ and $V\to\infty$ the Green functions depend on momenta only.
At $T\neq 0$ they depend        also on $T$.
The $\Gamma^{(2)}$-function is the inverse propagator:
$G^{(2)}\cdot\Gamma^{(2)}=1$. The observable corresponding to the
propagator is the space (equal-time) two-point correlation function
\be \kappa(|\vec x|)={T\over(2\pi)^3}\sum_{p_0}\int{d^3pe^{-i\vec
p\vec x}\over \Gamma^{(2)}(p_0^2,\vec p\,^2,T)}.  \ee The correlation
length $x_c(T)$ is determined by the nearest to the real axes zero of
the denominator of integrand (2) in the complex plane $|\,\vec p\,|$ at
$p_0=0$.  $$
\Gamma^{(2)}(0,\,-x_c^{-2}(T),\, T)=0.
$$ At $T=0$ the correlation length is simply the inverse mass of
the boson and at $T\neq0$ it defines the Debye screening.

The goal of this paper is to separate the vacuum (quantum)
fluctuations from thermal ones. Let us remind the well known one-loop
result for $\Gamma^{(2)}$-function
\be
\Gamma^{(2)}(p_0^2,\vec
p\,^2,T)=p^2+m^2_0+{\lambda\over2}\,{1\over (2\pi)^4}\int {d^4q\over
q^2+m^2}+{\lambda\over(2\pi)^3}\int{d^3q\over2\omega
(e^{\beta\omega}-1)},
\ee
where
$$
p^2=p_0^2+\vec p\,^2, \qquad \omega^2=\vec p\,^2+m^2.
$$
This expression clarifies what  we mean in ``vacuum'' and
``thermal'' fluctuations. We call the fist integral in (3)
``vacuum loop'' and the second one -- ``thermal loop''. The full loop
is divided in two terms -- additively in the first order of
perturbation theory
$$
\unitlength=2.00mm
\special{em:linewidth 0.6pt}
\linethickness{1.0pt}
\centering
\begin{picture}(52.00,4.90)
\emline{10.00}{1.00}{1}{18.00}{1.00}{2}
\put(14.00,1.00){\circle*{1.00}}
\put(14.00,3.00){\circle{3.80}}
\emline{26.00}{1.00}{3}{34.00}{1.00}{4}
\put(30.00,1.00){\circle*{1.00}}
\put(30.00,3.00){\circle{3.80}}
\emline{42.00}{1.00}{5}{50.00}{1.00}{6}
\put(46.00,1.00){\circle*{1.00}}
\put(46.00,3.00){\circle{3.80}}
\put(22.00,1.00){\makebox(0,0)[cc]{=}}
\put(38.00,1.00){\makebox(0,0)[cc]{+}}
\put(30.00,3.00){\makebox(0,0)[cc]{V}}
\put(46.00,3.00){\makebox(0,0)[cc]{T}}
\put(52.00,1.00){\makebox(0,0)[cc]{.}}
\end{picture}
$$
Thermal fluctuations
exponentially vanish at $T\to 0$ and
$$
\Gamma^{(2)}(p_0^2,\vec
p\,^2,T\equiv0)=\Gamma^{(2)}(p^2)
$$
where $\Gamma^{(2)}(p^2)$ is the usual
inverse propagator. Let us define the temperature dependent part by
extracting from the full propagator its value at zero temperature
\be
\gamma(p_0^2,\vec p\,^2,T)=\Gamma^{(2)}(p_0^2,\vec p\,^2,T)-
\Gamma^{(2)}(p^2).
\ee
Since the $2\to2$ scattering amplitude in the 1st
order of $PT$ is
$$
A^{(2)}(p,q;\,p',q')=-\lambda
$$
the $\gamma$-function can be written as
\be
\gamma(p_0,2,\vec
p\,^2,T)=-{1\over(2\pi)^3}\int{d^3qA^{(2)}(p,q;\,p,q)\over
2\omega(e^{\beta\omega}-1)}
\ee
Of course, this expression seems artificial for the trivial
1st order of $PT$ but it shows what  we search for higher orders.

\section{The second order of $PT$}

The calculations at $T\neq0$ are more complicated in 2nd order
compared to that at $T=0$ by two reasons: the regularization (cut off)
is not symmetric in momentum space and there is a loop summation
instead of loop integration. The last obstacle is important because at
finite temperature there appears loop
divergences depending on $T$. They come from different diagrams. One
of them contains a simple sum and the other -- the double sum. So
care should be taken when transforming sums to integrals to cancel the
above mentioned divergences.

The $\Gamma^{(2)}$-function in the 2nd order is saturated by
following diagrams
\newpic{1}
$$
\unitlength=2.00mm
\special{em:linewidth 0.6pt}
\linethickness{1.0pt}
\begin{picture}(76.50,7.00)(5,37)
\put(20.00,40.00){\makebox(0,0)[cc]{$=$}}
\emline{22.00}{40.00}{1}{30.00}{40.00}{2}
\put(32.00,40.00){\makebox(0,0)[cc]{$+$}}
\put(35.00,40.20){\makebox(0,0)[cc]{$\displaystyle{\lambda_0\over2}$}}
\put(41.00,40.00){\circle*{1.00}}
\put(41.00,42.00){\circle{3.80}}
\emline{37.00}{40.00}{3}{45.00}{40.00}{4}
\put(47.00,40.00){\makebox(0,0)[cc]{$-$}}
\put(50.00,40.30){\makebox(0,0)[cc]{$\displaystyle{\lambda_0^2\over4}$}}
\put(56.00,40.00){\circle*{1.00}}
\emline{52.00}{40.00}{5}{60.00}{40.00}{6}
\put(62.00,40.00){\makebox(0,0)[cc]{$-$}}
\put(65.00,40.30){\makebox(0,0)[cc]{$\displaystyle{\lambda_0^2\over6}$}}
\put(71.00,40.00){\circle{4.00}}
\emline{10.00}{40.00}{7}{6.00}{40.00}{8}
\emline{18.00}{40.00}{9}{14.00}{40.00}{10}
\put(69.00,40.00){\circle*{1.00}}
\put(73.00,40.00){\circle*{1.00}}
\put(56.00,41.35){\circle{2.70}}
\put(56.00,44.05){\circle{2.70}}
\put(56.00,42.70){\circle*{1.00}}
\put(12.00,40.00){\circle{4.00}}
\emline{10.18}{38.87}{11}{13.80}{40.94}{12}
\emline{9.89}{39.43}{13}{13.42}{41.47}{14}
\emline{12.36}{41.47}{15}{12.36}{41.47}{16}
\emline{10.58}{38.37}{17}{13.98}{40.30}{18}
\emline{11.25}{37.99}{19}{14.03}{39.56}{20}
\emline{9.86}{40.15}{21}{12.84}{41.87}{22}
\emline{10.16}{41.07}{23}{11.94}{42.08}{24}
\emline{12.49}{37.97}{25}{13.45}{38.55}{26}
\emline{67.00}{40.00}{27}{75.00}{40.00}{28}
\end{picture}\eqno(6)
$$

\noindent
and 1PI $\Gamma^{(4)}$-function

\newpic{2}
$$
\unitlength=2.00mm
\special{em:linewidth 0.6pt}
\linethickness{1.0pt}
\begin{picture}(77.50,8.00)(5,36)
\put(12.00,40.00){\circle{4.00}}
\emline{10.18}{38.87}{1}{13.80}{40.94}{2}
\emline{9.89}{39.43}{3}{13.42}{41.47}{4}
\emline{12.36}{41.47}{5}{12.36}{41.47}{6}
\emline{10.58}{38.37}{7}{13.98}{40.30}{8}
\emline{11.25}{37.99}{9}{14.03}{39.56}{10}
\emline{9.86}{40.15}{11}{12.84}{41.87}{12}
\emline{10.16}{41.07}{13}{11.94}{42.08}{14}
\emline{12.49}{37.97}{15}{13.45}{38.55}{16}
\put(19.00,40.00){\makebox(0,0)[cc]{$=\ -\lambda_0$}}
\put(27.00,40.00){\circle*{0.80}}
\put(34.00,40.00){\makebox(0,0)[cc]{$+\displaystyle{\lambda_0^2\over2}\
\biggl($}}
\put(49.00,40.00){\makebox(0,0)[cc]{$+$}}
\emline{10.49}{41.38}{17}{8.49}{43.45}{18}
\emline{13.51}{41.38}{19}{15.47}{43.49}{20}
\emline{8.57}{36.43}{21}{10.53}{38.55}{22}
\emline{15.43}{36.43}{23}{13.43}{38.51}{24}
\put(42.00,38.00){\oval(6.00,6.00)[t]}
\put(42.00,42.00){\oval(6.00,6.00)[b]}
\emline{24.00}{37.00}{25}{30.00}{43.00}{26}
\emline{24.00}{43.00}{27}{30.00}{37.00}{28}
\put(58.00,40.00){\oval(6.00,6.00)[l]}
\put(54.00,40.00){\oval(6.00,6.00)[r]}
\put(61.00,40.00){\makebox(0,0)[cc]{$+$}}
\put(39.50,40.00){\circle*{0.80}}
\put(44.50,40.00){\circle*{0.80}}
\put(56.00,42.17){\circle*{0.80}}
\put(56.00,37.67){\circle*{0.80}}
\put(66.50,40.00){\oval(5.00,6.00)[r]}
\put(70.50,40.00){\oval(5.00,4.00)[l]}
\put(68.83,41.50){\circle*{0.80}}
\put(68.83,38.33){\circle*{0.80}}
\emline{70.52}{42.01}{29}{71.26}{41.75}{30}
\emline{70.52}{38.00}{31}{71.33}{38.35}{32}
\emline{71.26}{41.73}{33}{75.10}{38.00}{34}
\emline{71.34}{38.36}{35}{72.66}{39.67}{36}
\emline{72.66}{39.67}{37}{72.80}{39.62}{38}
\emline{72.80}{39.62}{39}{73.00}{39.58}{40}
\emline{73.00}{39.58}{41}{73.16}{39.61}{42}
\emline{73.16}{39.61}{43}{73.30}{39.69}{44}
\emline{73.30}{39.69}{45}{73.45}{39.85}{46}
\emline{73.45}{39.85}{47}{73.49}{39.98}{48}
\emline{73.49}{39.98}{49}{73.45}{40.17}{50}
\emline{73.45}{40.17}{51}{73.41}{40.30}{52}
\emline{73.41}{40.30}{53}{75.12}{42.00}{54}
\put(78.00,40.00){\makebox(0,0)[cc]{$\biggr).$}}
\end{picture}
\eqno(7)
$$

The appropriate regularization is
$|\,\vec p\,|\leq\Lambda$, i.e. it is a cylinder in momentum space
rather then a sphere as usual at $T=0$. Loop summations over zero's
components of momenta are convergent by itself.  Subtraction points
are defined at $T=0$ as follows
$$
\Gamma^{(2)}(p^2)\biggm|_{p^2=-m}=0,\qquad {\partial\Gamma
^{(2)}(p^2)\over\partial p^2}\biggm|_{p^2=0}=1,\eqno(8)
$$
$$
\Gamma^{(4)}(p_1,p_2,p_3,p_4)\biggm|_{p_i=0}=-\lambda,\eqno(9)
$$
where $m$ and $\lambda$ are renormalized (observable), mass and
coupling constant.

The technical problem is to transform the loop sums into
integrals. For the simple sum over $q_0$ the answer is [12]
$$
S\nu=T\sum_{q_0}{1\over(q^2+m^2)^\nu}={1\over2\pi}\int{dq_0\over
(q^2+m^2)^\nu}\left(1+{2\over e^{-i\beta q_0}-1}\right),
\nu=1,2,
\eqno(10)
$$
or
$$
S_1={1\over2\pi}\int{dq_0\over Q^2+m^2}+{1\over
\omega(e^{\beta\omega}-1)},\eqno(11)
$$
$$
S_2={1\over2\pi}\int{dq_0\over Q^2+m^2}+{1\over
\omega}\,{\partial\over\partial\omega}\left({1\over
2\omega(e^{\beta\omega}-1)}\right).\eqno(12)
$$
There is more trouble with double sum $S_3$ contained in the sunrise
diagram\ \
\newpic{3}
$
\unitlength=1.00mm
\special{em:linewidth 0.6pt}
\linethickness{1.0pt}
\begin{picture}(10.00,4.00)(8,39)
\put(14.00,40.00){\circle{4.00}}
\put(12.00,40.00){\circle*{1.00}}
\put(16.00,40.00){\circle*{1.00}}
\emline{8.00}{40.00}{1}{20.00}{40.00}{2}
\end{picture}$\ \ ,

$$
S_3=T^2\sum_{q_0k_0}{1\over(q^2+m^2)(k^2+m^2)[(q+k-p)^2+m^2]}.\eqno(13)
$$
The substitution (10) is not correct in this case owing to the pole
surface in the complex manifold $q_0\otimes k_0$ caused by the factor
$[(q+k-p)^2+m^2]^{-1}$ in (13). Careful manipulations give the
result
\bea
&&S_3={1\over(2\pi)^2}\int{dq_0dk_0\over(q^2+m^2)
(k^2+m^2)[(q+k-p)^2+m^2]}+ \nonumber\\
&&+{3\over2\omega_k(e^{\beta\omega_k}-1)}{1\over2\pi}\int{dq_0\over(q^2+m^2)}
\left({1\over(q+k-p)^2+m^2}+{1\over(q-k-p)^2+m^2}\right)+ \nonumber\\
&&+{3\over2\omega_k(e^{\beta\omega_k}-1)2\omega_q(e^{\beta\omega_q}-1)}\times
\left({1\over(k+q+p)^2+m^2}+\right.\nonumber\\
&&+\left.{1\over(k-q+p)^2+m^2}+
{1\over(q-k+p)^2+m^2}+{1\over(k+q-p)^2+m^2}\right)\ \ \ \ \ \ \ \ \ \
(14) \nonumber
\eea
The second diagram in the r.h.s. of (6) is renormalized in the 2nd
order of $PT$ with the help of eqs. (6)--(9) i.e.
$$
\lambda_0=\lambda+{3\lambda^2\over2}{1\over(2\pi)^4}\int{d^4q\over
(q^2+m^2)^2},
$$
$$
m_0^2=m^2-{\lambda\over2}{1\over(2\pi)^4}\int{d^4q\over q^2+m^2}.
$$
Therefore up to the power $\lambda^2$ we have
\newpic{4}
$$
\unitlength=2mm
\special{em:linewidth 0.6pt}
\linethickness{1.0pt}
\begin{picture}(23.00,4.00)(35,38)
\put(10.00,40.20){\makebox(0,0)[cc]{$\displaystyle{\lambda_0\over2}$}}
\put(16.00,40.00){\circle*{1.00}}
\put(16.00,41.50){\circle{2.90}}
\emline{12.00}{40.00}{1}{20.00}{40.00}{2}
\put(23.00,40.00){\makebox(0,0)[lc]{$=\displaystyle{\frac{\lambda_0}2T\sum
\limits_{q_0}\frac1{(2\pi)^3}\int\frac{d^3q}{q^2+m_0^2}=
\frac{\lambda}2T\sum\limits_{q_\nu}\frac1{(2\pi)^3}\int\frac{d^3q}{q^2+m^2}+}$}}
\end{picture}
$$
$$
\quad\quad\quad+{\lambda^2\over4}\biggl(T\sum_{q_0}{1\over(2\pi)^3}\int
{d^3q\over(q^2+m^2)^2}\biggr)\cdot\biggl({1\over(2\pi)^4}\int{d^4k\over
k^2+m^2}\biggr)+\eqno(15)
$$
$$\quad\quad\quad+{3\lambda^2\over4}\biggl({1\over(2\pi)^4}\int{d^4q\over(q^2+m^2)^2}
\biggr)\cdot\biggl(T{\sum_{k_0}}{1\over(2\pi)^3}\int{d^3k\over
k^2+m^2}\biggr).
$$
\newpic{5}
$$
{\unitlength=2.00mm
\special{em:linewidth 0.6pt}
\linethickness{1.0pt}
\begin{picture}(20.00,15.40)(35,0)
\put(10.00,10.30){\makebox(0,0)[cc]{$\displaystyle{\lambda_0^2\over4}$}}
\put(16.00,10.00){\circle*{1.00}}
\emline{12.00}{10.00}{1}{20.00}{10.00}{2}
\put(16.00,11.35){\circle{2.70}}
\put(16.00,14.05){\circle{2.70}}
\put(16.00,12.70){\circle*{1.00}}
\put(23.00,10.00){\makebox(0,0)[lc]{$\displaystyle{
={\lambda^2\over4}\biggl(T\sum\limits_{q_0}{1\over(2\pi)^3}\int
{d^3q\over(q^2+m^2)^2}\biggr)
\biggl(T\sum\limits_{k_0}{1\over(2\pi)^3}\int{
d^3k\over k^2+m^2}\biggr).}\ \quad\quad(16)$}}
\end{picture}}
$$

\newpage
\newpic{6}
$$
\unitlength=2.00mm
\special{em:linewidth 0.6pt}
\linethickness{1.0pt}
\begin{picture}(20.00,12.00)(35,0)
\put(10.00,10.30){\makebox(0,0)[cc]{$\displaystyle{\lambda_0^2\over6}$}}
\put(16.00,10.00){\circle{2.70}}
\put(14.65,10.00){\circle*{0.80}}
\put(17.35,10.00){\circle*{0.80}}
\emline{12.00}{10.00}{1}{20.00}{10.00}{2}
\put(23.00,10.00){\makebox(0,0)[lc]{$\displaystyle{
={\lambda^2\over6}T^2
\sum\limits_{q_0k_0}{1\over(2\pi)^6}\int{d^3q\,
d^3k\over (q^2+m^2)(k^2+m^2)[(q+k-p)^2+m^2]}}\ \quad\quad(17)$}}
\end{picture}
$$
Let us write the temperature dependent part of $\Gamma^{(2)}$-function
(4) as a sum of two terms
$$
\gamma(p_0^2,\vec p\,^2,T)=\gamma_2(p_0^2,\vec p\,^2,T)+
\gamma_3(p_0^2,\vec p\,^2,T).
$$
Then combining (15)--(17) and regarding transformations from sums to integrals
 (10)--(14) we obtain
$$
\gamma_2(p_0^2,\vec p\,^2,T)=-{1\over(2\pi)^3}\int{
d^3q\over
2\omega_q(e^{\beta\omega_q}-1)}\left\{-\lambda+{\lambda^2\over2}
{1\over(2\pi)^4}\int{d^4k\over k^2+m^2}\times\right.
$$
$$
\times\left[{1\over(k-q-p)^2+m^2}+{1\over (k+q-p)^2+m^2}-{2\over
k^2+m^2}\right]\biggl.\biggr\},\eqno(18)
$$
where $q_0=i\omega_q$;
$$\gamma_3(p_0^2,\vec
p\,^2,T)=-{1\over2!(2\pi)^6}\int{d^3q\, d^3k\over
2\omega_q(e^{\beta\omega_q}-1)2\omega_k(e^{\beta\omega_k}-1)}
\lambda^2
\left[{1\over(k+q+p)^2+m^2}+  \right.
$$
$$
+{1\over(k+q-p)^2+m^2} +
\left.{1\over (k+p-q)^2+m^2}+{1\over(q+p-k)^2+m^2}+{1\over \vec
q\,^2}\right],\eqno(19)$$
where $k_0=i\omega_k$, $q_0=i\omega_q$.

\noindent
One  can recognize in the braces of (18) the renormalized $2\to 2$
scattering amplitude at zero angle in the 2nd order of $PT$. So
expected expression (5) is confirmed in 2nd order of $PT$.

\noindent
Now we assume by analogy that $\gamma_3$ should have the form
$$
\gamma_3(p_0^2,\vec p\,^2,T)=-{1\over2!(2\pi)^6}\int{d^3q\,
d^3k\,
A^{(3)}(p,q,k;\,p,q,k)\over2\omega_q(e^{\beta\omega_q}-1)
2\omega_k(e^{\beta\omega_k}-1)}\eqno(20)
$$
Really, the lowest order in which the $3\to 3$ scattering amplitude
appears is $\lambda^2$. There are 10 tree diagrams (the direct channel
diagram plus 9 exchanged diagrams) which contribute to the $3\to 3$
amplitude. Collecting them in 3 subsets as shown below

\newpage
\begin{figure}[h]
\unitlength=0.7mm
\special{em:linewidth 0.6pt}
\linethickness{1.0pt}
\begin{picture}(189.00,100.00)(-5,-38)
\emline{0.00}{40.00}{1}{30.00}{40.00}{2}
\emline{10.00}{40.00}{3}{0.00}{50.00}{4}
\emline{10.00}{40.00}{5}{0.00}{30.00}{6}
\put(10.00,40.00){\circle*{2.00}}
\emline{20.00}{40.00}{7}{30.00}{50.00}{8}
\emline{20.00}{40.00}{9}{30.00}{30.00}{10}
\put(20.00,40.00){\circle*{2.00}}
\put(-3.00,50.00){\makebox(0,0)[cc]{$k$}}
\put(-3.00,40.00){\makebox(0,0)[cc]{$q$}}
\put(-3.00,30.00){\makebox(0,0)[cc]{$p$}}
\put(33.00,50.00){\makebox(0,0)[cc]{$k'$}}
\put(33.00,40.00){\makebox(0,0)[cc]{$q'$}}
\put(33.00,30.00){\makebox(0,0)[cc]{$p'$}}
\emline{65.00}{47.50}{11}{80.00}{47.50}{12}
\emline{65.00}{47.50}{13}{50.00}{52.50}{14}
\emline{65.00}{47.50}{15}{50.00}{42.50}{16}
\emline{50.00}{32.50}{17}{65.00}{32.50}{18}
\emline{65.00}{32.50}{19}{80.00}{37.50}{20}
\emline{65.00}{32.50}{21}{80.00}{27.50}{22}
\emline{65.00}{47.50}{23}{65.00}{32.50}{24}
\put(65.00,32.50){\circle*{2.00}}
\put(65.00,47.50){\circle*{2.00}}
\put(47.00,52.50){\makebox(0,0)[cc]{$k$}}
\put(47.00,42.50){\makebox(0,0)[cc]{$q$}}
\put(47.00,32.50){\makebox(0,0)[cc]{$p$}}
\put(83.00,47.50){\makebox(0,0)[cc]{$p'$}}
\put(83.00,37.50){\makebox(0,0)[cc]{$k'$}}
\put(83.00,27.50){\makebox(0,0)[cc]{$q'$}}
\emline{100.00}{47.50}{25}{115.00}{47.50}{26}
\emline{115.00}{47.50}{27}{130.00}{52.50}{28}
\emline{115.00}{47.50}{29}{130.00}{42.50}{30}
\put(115.00,47.50){\circle*{2.00}}
\put(97.00,47.50){\makebox(0,0)[cc]{$q$}}
\put(133.00,52.50){\makebox(0,0)[cc]{$p'$}}
\put(133.00,42.50){\makebox(0,0)[cc]{$k'$}}
\emline{115.00}{32.50}{31}{130.00}{32.50}{32}
\emline{115.00}{32.50}{33}{100.00}{37.50}{34}
\emline{115.00}{32.50}{35}{100.00}{27.50}{36}
\put(115.00,32.50){\circle*{2.00}}
\put(97.00,37.50){\makebox(0,0)[cc]{$k$}}
\put(97.00,27.50){\makebox(0,0)[cc]{$p$}}
\put(133.00,32.50){\makebox(0,0)[cc]{$q'$}}
\emline{115.00}{47.50}{37}{115.00}{32.50}{38}
\emline{150.00}{47.50}{39}{165.00}{47.50}{40}
\emline{165.00}{47.50}{41}{180.00}{52.50}{42}
\emline{165.00}{47.50}{43}{180.00}{42.50}{44}
\put(165.00,47.50){\circle*{2.00}}
\put(147.00,47.50){\makebox(0,0)[cc]{$k$}}
\put(183.00,52.50){\makebox(0,0)[cc]{$p'$}}
\put(183.00,42.50){\makebox(0,0)[cc]{$q'$}}
\emline{165.00}{32.50}{45}{180.00}{32.50}{46}
\emline{165.00}{32.50}{47}{150.00}{37.50}{48}
\emline{165.00}{32.50}{49}{150.00}{27.50}{50}
\put(165.00,32.50){\circle*{2.00}}
\put(147.00,37.50){\makebox(0,0)[cc]{$q$}}
\put(147.00,27.50){\makebox(0,0)[cc]{$p$}}
\put(183.00,32.50){\makebox(0,0)[cc]{$k'$}}
\emline{165.00}{47.50}{51}{165.00}{32.50}{52}
\emline{0.00}{16.50}{53}{15.00}{16.50}{54}
\emline{15.00}{16.50}{55}{30.00}{21.50}{56}
\emline{15.00}{16.50}{57}{30.00}{11.50}{58}
\put(15.00,16.50){\circle*{2.00}}
\put(-3.00,16.50){\makebox(0,0)[cc]{$k$}}
\put(33.00,21.50){\makebox(0,0)[cc]{$k'$}}
\put(33.00,11.50){\makebox(0,0)[cc]{$q'$}}
\emline{15.00}{1.50}{59}{30.00}{1.50}{60}
\emline{15.00}{1.50}{61}{0.00}{6.50}{62}
\emline{15.00}{1.50}{63}{0.00}{-3.50}{64}
\put(15.00,1.50){\circle*{2.00}}
\put(-3.00,6.50){\makebox(0,0)[cc]{$q$}}
\put(-3.00,-3.50){\makebox(0,0)[cc]{$p$}}
\put(33.00,1.50){\makebox(0,0)[cc]{$p'$}}
\emline{15.00}{16.50}{65}{15.00}{1.50}{66}
\emline{65.00}{16.50}{67}{80.00}{16.50}{68}
\emline{65.00}{16.50}{69}{50.00}{21.50}{70}
\emline{65.00}{16.50}{71}{50.00}{11.50}{72}
\emline{50.00}{1.50}{73}{65.00}{1.50}{74}
\emline{65.00}{1.50}{75}{80.00}{6.50}{76}
\emline{65.00}{1.50}{77}{80.00}{-3.50}{78}
\emline{65.00}{16.50}{79}{65.00}{1.50}{80}
\put(65.00,1.50){\circle*{2.00}}
\put(65.00,16.50){\circle*{2.00}}
\put(47.00,21.50){\makebox(0,0)[cc]{$k$}}
\put(47.00,11.50){\makebox(0,0)[cc]{$q$}}
\put(47.00,1.50){\makebox(0,0)[cc]{$p$}}
\put(83.00,16.50){\makebox(0,0)[cc]{$k'$}}
\put(83.00,6.50){\makebox(0,0)[cc]{$q'$}}
\put(83.00,-3.50){\makebox(0,0)[cc]{$p'$}}
\emline{100.00}{16.50}{81}{115.00}{16.50}{82}
\emline{115.00}{16.50}{83}{130.00}{21.50}{84}
\emline{115.00}{16.50}{85}{130.00}{11.50}{86}
\put(115.00,16.50){\circle*{2.00}}
\put(97.00,16.50){\makebox(0,0)[cc]{$q$}}
\put(133.00,21.50){\makebox(0,0)[cc]{$q'$}}
\put(133.00,11.50){\makebox(0,0)[cc]{$k'$}}
\emline{115.00}{1.50}{87}{130.00}{1.50}{88}
\emline{115.00}{1.50}{89}{100.00}{6.50}{90}
\emline{115.00}{1.50}{91}{100.00}{-3.50}{92}
\put(115.00,1.50){\circle*{2.00}}
\put(97.00,6.50){\makebox(0,0)[cc]{$k$}}
\put(97.00,-3.50){\makebox(0,0)[cc]{$p$}}
\put(133.00,1.50){\makebox(0,0)[cc]{$p'$}}
\emline{115.00}{16.50}{93}{115.00}{1.50}{94}
\emline{165.00}{16.50}{95}{180.00}{16.50}{96}
\emline{165.00}{16.50}{97}{150.00}{21.50}{98}
\emline{165.00}{16.50}{99}{150.00}{11.50}{100}
\emline{150.00}{1.50}{101}{165.00}{1.50}{102}
\emline{165.00}{1.50}{103}{180.00}{6.50}{104}
\emline{165.00}{1.50}{105}{180.00}{-3.50}{106}
\emline{165.00}{16.50}{107}{165.00}{1.50}{108}
\put(165.00,1.50){\circle*{2.00}}
\put(165.00,16.50){\circle*{2.00}}
\put(147.00,21.50){\makebox(0,0)[cc]{$q$}}
\put(147.00,11.50){\makebox(0,0)[cc]{$k$}}
\put(147.00,1.50){\makebox(0,0)[cc]{$p$}}
\put(183.00,16.50){\makebox(0,0)[cc]{$q'$}}
\put(183.00,6.50){\makebox(0,0)[cc]{$k'$}}
\put(183.00,-3.50){\makebox(0,0)[cc]{$p'$}}
\emline{50.00}{-17.50}{109}{65.00}{-17.50}{110}
\emline{65.00}{-17.50}{111}{80.00}{-12.50}{112}
\emline{65.00}{-17.50}{113}{80.00}{-22.50}{114}
\put(65.00,-17.50){\circle*{2.00}}
\put(47.00,-17.50){\makebox(0,0)[cc]{$k$}}
\put(83.00,-12.50){\makebox(0,0)[cc]{$p'$}}
\put(83.00,-22.50){\makebox(0,0)[cc]{$k'$}}
\emline{65.00}{-32.50}{115}{80.00}{-32.50}{116}
\emline{65.00}{-32.50}{117}{50.00}{-27.50}{118}
\emline{65.00}{-32.50}{119}{50.00}{-37.50}{120}
\put(65.00,-32.50){\circle*{2.00}}
\put(47.00,-27.50){\makebox(0,0)[cc]{$q$}}
\put(47.00,-37.50){\makebox(0,0)[cc]{$p$}}
\put(83.00,-32.50){\makebox(0,0)[cc]{$q'$}}
\emline{65.00}{-17.50}{121}{65.00}{-32.50}{122}
\emline{100.00}{-17.50}{123}{115.00}{-17.50}{124}
\emline{115.00}{-17.50}{125}{130.00}{-12.50}{126}
\emline{115.00}{-17.50}{127}{130.00}{-22.50}{128}
\put(115.00,-17.50){\circle*{2.00}}
\put(97.00,-17.50){\makebox(0,0)[cc]{$q$}}
\put(133.00,-12.50){\makebox(0,0)[cc]{$p'$}}
\put(133.00,-22.50){\makebox(0,0)[cc]{$q'$}}
\emline{115.00}{-32.50}{129}{130.00}{-32.50}{130}
\emline{115.00}{-32.50}{131}{100.00}{-27.50}{132}
\emline{115.00}{-32.50}{133}{100.00}{-37.50}{134}
\put(115.00,-32.50){\circle*{2.00}}
\put(97.00,-27.50){\makebox(0,0)[cc]{$k$}}
\put(97.00,-37.50){\makebox(0,0)[cc]{$p$}}
\put(133.00,-32.50){\makebox(0,0)[cc]{$k'$}}
\emline{115.00}{-17.50}{135}{115.00}{-32.50}{136}
\put(210.00,40.00){\makebox(0,0)[cc]{(21)}}
\put(210.00,11.50){\makebox(0,0)[cc]{(22)}}
\put(210.00,-27.50){\makebox(0,0)[cc]{(23)}}
\end{picture}
\end{figure}

\noindent
one can see that the first four terms in square brackets of (19) are
just the four diagrams (21) with the allowance $p=p'$, $q=q'$,
$k=k'$ (zero angles scattering). More problematic are
 the diagrams (22). On mass shell of momenta $q$ or $k$ these
diagrams diverge themselves and should be regularized. Several
papers were devoted to this problem [13]. Our direct field-theory
calculations give the simple recipe for such regularization. Namely
one must set $p-p'=\epsilon=(\epsilon_o,\vec \epsilon\,)$ and take
the limit in the final result in the following order

$$
\gamma_3(p_0^2,\vec p\,^2,T)=\lim_{\vec
\epsilon=0}\left(\lim_{\epsilon_0=0}{-1\over2!(2\pi)^3}\int{d^3q\,
d^3k\,
A^{(3)}(p,q,k;\,p+\epsilon,q,k)\over2\omega_q(e^{\beta\omega_q}-1)
2\omega_k(e^{\beta\omega_k}-1)}\right).
$$
As to the last two diagrams (23), they do not contribute to
$\gamma_3$ because the considered
$\Gamma^{(2)}$-function by definition should contain 1PI diagrams
only.

\section{Conclusion}

Surprisingly the temperature dependent part of
$\Gamma^{(2)}(p_0^2,\vec p\,^2,T)$ -- eqs. (5) and (20) -- can be
expressed in an extremely compact and physically transparent form
through the scattering amplitudes.  These low order $PT$ calculations
give us the opportunity to generalize naturally enough the obtained
result in the following way
$$
\gamma(p_0^2,\vec
p\,^2,T)=\sum_{n=2}^{\infty}\gamma_n(p_0^2,\vec p\,^2,T),
$$
$$\gamma_{n+1}(p_0^2,\vec p\,^2,T)=\eqno(24)$$
$$=-{1\over n!(2\pi)^{3n}}\int
\left(\prod^n_{l=1}{d^3q_l\over2\omega_l(e^{\beta\omega_l}-1)}\right)
A^{(n+1)}
(p,q_1,\dots,q_n;\, p,q_1,\dots,q_n).$$
This expression is linear in the scattering amplitudes and looks like a
contribution of so called ``first-type'' diagrams in the $S$-matrix
formulation of statistical mechanics. We can not exclude without
further analysis the appearence of some bilinear terms in amplitudes
 in the exact formula for $\gamma(p_0^2,\vec p\,^2,T)$ similar to
the so-called ``second-type'' diagrams. But we shall not be
surprised if the eq.  (24) satisfies the exact representation.

We thank profs. F.~Paccanoni, B.V.~Struminsky and E.S.Martynov
 for discussions.

\end{document}